\documentclass[prl,aps,twocolumn,showpacs,preprintnumbers,floats,amsmath,amssymb]{revtex4}

\usepackage{graphicx}
\usepackage{dcolumn}
\usepackage{bm}

\def \be{\begin{equation}}
\def \ee{\end{equation}}

\begin{document}

\title{Current noise through a Kondo quantum dot in a SU(N) Fermi liquid state}

\author{Christophe Mora$^1$}
\author{Xavier Leyronas$^2$}
\author{Nicolas Regnault$^1$}
\affiliation{$^1$~Laboratoire Pierre Aigrain, ENS, Universit\'e Denis Diderot 7, CNRS; 24 rue Lhomond, 75005 Paris, France}
\affiliation{$^2$~Laboratoire de Physique Statistique, D\'epartement de Physique,
Ecole Normale Sup\'erieure,
24 rue Lhomond, F-75005 Paris, France}

\date{\today}

\begin{abstract} 
The current noise through a mesoscopic quantum dot is calculated and 
analyzed in the Fermi liquid regime of the SU(N) Kondo model.
Results connect the Johnson-Nyquist noise to the shot noise for an arbitrary
ratio of voltage and temperature, and show that temperature
corrections are sizeable in usual experiments.
For the experimentally relevant SU(4) case, quasiparticle interactions are shown to increase the shot noise.

\end{abstract}

\pacs{72.70.+m, 72.15.Qm, 71.10.Ay, 73.63.Fg, 73.63.Kv}
\maketitle

The Kondo effect~\cite{hewson1993}, {\it i.e.} the screening of a local spin
by coupling to conduction electrons, is a paradigm for strongly correlated systems 
as it exhibits sophisticated many-body correlations with a quite simple model.
Its tunable realization in mesoscopic quantum dots, semiconductors, or carbon 
nanotubes, has triggered a renewed interest~\cite{kouwen2001+glazman2005} in Kondo physics 
probed by transport measurements. 
In the ground state, the dot spin is screened by a cloud of conduction electrons
and forms a singlet. Low energy properties of remaining electrons 
are described by a local Fermi liquid theory~\cite{nozieres1974}. 
In this theory, electrons are 
scattered elastically by the singlet in a similar way as by a resonant level.
Electrons also interact through polarization of the spin singlet. 
The ratio of elastic to inelastic scattering is fixed by universality, {\it i.e.}
the Kondo temperature $T_K$ is the only scale that governs low energy properties
of the model. 
This description applies to SU(2) symmetry but also more generally
to SU(N). In that case, both the spin and orbital degrees
of freedom are screened by delocalized electrons in the reservoirs.
In particular, the SU(4) case
has become recently the subject of extensive investigation. Various experimental
settings have been proposed~\cite{borda2003+lehur2003a+lehur2003b+zarand2003,choi2005+eto2005}
and its realization has been reported already in vertical quantum dots~\cite{sasaki2004}
and carbon nanotubes~\cite{herrero2005+makarovski}. 

A promising experimental tool to study Kondo physics is current noise measurement.
These experiments are technically challenging in
the Kondo regime notwithstanding recent progresses~\cite{onac2006+wu2007+herrmann2007}.
Noise can probe out of equilibrium properties of the model.
 This has not been much
investigated so far since only a few theoretical methods~\cite{meir2002,golub2006,gogolin2006a} 
apply to the out of equilibrium
situation in comparison with the equilibrium case. 
In particular, the shot-noise at low temperature could provide information on the
statistics of charge transfer.
Interestingly, a picture has emerged recently~\cite{sela2006,gogolin2006b} for 
the SU(2) Kondo effect at low energy where scattering
events of two electrons lead to an effective charge of $5/3 \,e$ in the backscattering current.
This emphasizes the strong role of interactions for charge transfer
even in the vicinity of a Fermi liquid fixed point.

The heating due to voltage polarization and the decoupling of phonons to electrons 
at low energy implies that the  
temperature of electrons is never really small in practical situations.
It is therefore highly
desirable to have a theory that holds at finite temperature.
The purpose of this letter is to provide a general analysis for the zero-frequency 
current noise $S \equiv 2 \int dt \langle \Delta \hat{I}(t) \Delta \hat{I}(0) \rangle $
in the SU(N) Fermi liquid regime. $\hat{I}$ is the current operator. 
Our expressions are valid at low temperature $T$
and voltage $V$ and for any value of $V/T$. They interpolate between the shot-noise $T\ll e V$
and the Johnson-Nyquist $e V \ll T$ limits where the fluctuation-dissipation relation
to the conductance is recovered.
Note that the zero-temperature limit has been addressed recently~\cite{vitu2007}.
Three uncorrelated processes were identified and interpreted for the shot-noise ($T \ll e V$).
The first two are backscattering of charges $e (2 {\cal T}_0-1)$ and $2 e (2 {\cal T}_0-1)$, ${\cal T}_0$ being
the transmission at vanishing energy.
They give the charge $5/3 \,e$ derived in Refs.~\cite{sela2006,gogolin2006b} for SU(2).
The third process increases available electron-hole pairs leading to an enhancement of the partition noise.

We consider a quantum dot attached to two leads with source and drain. Only a single,
possibly degenerate, energy level participates to transport.
While the dot spin has usually SU(2) symmetry, a higher SU(N) symmetry can be achieved 
if orbital degeneracy of the dot level takes place, resulting in 
a larger N-component pseudospin.
Below the strong Coulomb energy $U$, a SU(N) Kondo model is obtained coupling 
antiferromagnetically the spin of the dot with lead electrons. 
We assume here that the dot has exactly one electron.
Below the Kondo temperature $T_K$ ($\ll U$), the dot spin forms a SU(N) singlet
with lead electrons. 
This requires however that any orbital index is conserved during lead-dot 
tunneling processes~\cite{choi2006}. Experiments currently 
investigate Kondo SU(4)~\cite{sasaki2004,herrero2005+makarovski}. 
In double-dot structure
the orbital index discriminates the two dots, in carbon nanotubes it originates from the
K-K' orbital degeneracy of the graphene band structure.

In the regime of energies (temperature and voltage) smaller than $T_K$, 
 the dot spin is frozen out in the singlet configuration
and  Fermi liquid theory applies to lead electrons~\cite{nozieres1974}.
We first detail elastic scattering. The 
low energy form of the dot Green's function~\cite{hewson1993} results in
an (s-wave) electron phase shift~\cite{langreth1966}, 
$\delta_{\rm el} (\epsilon) = \arctan[\Gamma/(\epsilon_K - \epsilon)]$, {\it i.e.} the same as for 
a resonant level of energy $\epsilon_K$ and half-width $\Gamma$, 
$\epsilon$ being measured from the Fermi level. Expanding to second order in $\epsilon$ leads to
\begin{equation}\label{dephasing}
\delta_{\rm el} (\epsilon) = \delta_0 + \frac{\alpha_1}{T_K} \epsilon 
+ \frac{\alpha_2}{T_K^2} \epsilon^2,
\end{equation}
with $\alpha_1/T_K = \sin^2 (\delta_0)/\Gamma$ and
$\delta_0\equiv \delta_{\rm el} (0)= \pi/N$, the phase shift at zero energy, is imposed
by the Friedel sum rule~\cite{langreth1966}.  
The value of $\alpha_1 \sim 1/N$ depends on the precise
definition of the Kondo temperature and is therefore not universal. In contrast, the ratio 
$\alpha_2/\alpha_1^2 = \cot \delta_0$ is universal.
For SU(2),  $\alpha_2 =0$, and there is no
 $\epsilon^2$ dependence in contrast to SU(N)  ($N> 2$) as noted by Ref.~\cite{vitu2007}.
Electron interaction -through polarization of the spin singlet- is written below 
on the basis of scattering states that include the
phase shift Eq.~\eqref{dephasing}.

Close to the SU(2) unitary limit, Ref.~\cite{kaminski2000} proposes 
to write the current as $\hat{I} = I_u - \hat{I}_{BS}$ with $I_u= 2 \frac{e^2}{h} V$ the
unitary contribution and $\hat{I}_{BS}$ the backscattering current. 
Current and shot noise can then be derived but not the noise at finite temperature 
since correlations between $I_u$ and $\hat{I}_{BS}$ are not included.
In particular, the Johnson-Nyquist relation to the conductance is not recovered.
We use here  a more general approach where the current operator
is expanded over a convenient basis of scattering states. Elastic scattering
is then easily described.
We separate the lead electron field into its symmetric and 
antisymmetric parts $\psi (x) = \psi_s (x) + \psi_a (x)$.
The $x$ axis is oriented from left (source) to right (drain).
We consider the case where the dot is coupled 
symmetrically to the two leads. In that case, only the symmetric part $\psi_s$ 
(s-wave)
is modified by the Kondo coupling~\cite{kouwen2001+glazman2005}. Therefore we can expand over eigenstates
of the free problem (we write $\sum_k \equiv \int \frac{d k}{2 \pi}$ throughout)
\begin{equation}\label{antisym}
\psi_a (x) = \sum_k  \frac{e^{i ( k_F +k) x} - e^{-i ( k_F +k) x} }{\sqrt{2}}   \, \, a_k,
\end{equation}
where $a_k$ annihilates an antisymmetric mode with energy $\epsilon_k = \hbar v_F k$.
For symmetric modes of energy $\epsilon_k$, electrons are reflected 
at $x=0$ with dephasing $2 \delta_{\rm el} (\epsilon_k)$, see Eq.~\eqref{dephasing}.
For $x<0$, they read  $( e^{i ( k_F +k) x} - e^{2 i \delta_{\rm el} (\epsilon_k)}
e^{-i ( k_F +k) x} )/\sqrt{2}$. Left (right) scattering states are obtained 
as (anti)symmetric combination of the symmetric and antisymmetric modes.
$\delta_{\rm el}=0$ (resp. $\delta_{\rm el} = \pi/2$) corresponds to totally reflected (resp. transmitted)
scattering states. Following Ref.~\cite{affleck1993}, we write the symmetric
part for $x<0$ as
\begin{equation}\label{sym}
\psi_s (x) = \frac{1}{\sqrt{2}} ( e^{i k_F x} b (x) - e^{-i k_F x} \tilde{b} (x) ).
\end{equation}
 $b(x)$ ($\tilde{b}(x)$) describes incoming (outgoing) waves. 
Writting  $b(x) = \sum_k b_k e^{i k x}$ defines an extension of $b(x)$ to $x>0$
which has nothing to do with the physical $x>0$ half-space. Physically,
it corresponds to unfolding the outgoing part of the incoming wave to the positive axis. 
The ${\cal S}$-matrix relates  $b(x)$ and $\tilde{b}(x)$ through the boundary condition
$\tilde{b}(x) = {\cal S} \, b(-x) \equiv \sum_k e^{2 i \delta_{\rm el} (\epsilon_k)} e^{- i k x} b_k$.
Collecting Eqs.~\eqref{antisym} and \eqref{sym}, we obtain a compact expression for
the symmetrized current passing through the dot
\begin{equation}\label{current}
\hat{I} = \frac{e}{2 \nu h} ( a^\dagger(x) b(x) - a^\dagger(-x) {\cal S} b(-x) + {\rm h.c.} ),
\end{equation}
with $a(x) \equiv \sum_k e^{i k x} a_k$ and arbitrary $x<0$.
$\nu = 1/ (h v_F)$ is the density of state for 1D fermions moving along one direction.
This expression can be generalized straightforwardly to spinfull fermions.

A dc bias is applied between electrodes, $\mu_L  = - \mu_R = eV /2$ where
symmetric capacitive coupling is assumed.
This defines the population of left and right scattering states with annihilation
operators $c_{L/R,k} = (b_k \pm a_k)/\sqrt{2}$
such that $\langle a_k^\dagger a_{k'} \rangle = \langle b_k^\dagger b_{k'} \rangle
=\delta(k-k') f_{aa} (k)$ and $\langle b_k^\dagger a_{k'} \rangle = \delta(k-k') f_{ab} (k)$
with the notations $f_{aa} (k) \equiv \frac{1}{2} \sum_{\pm}  f(\epsilon_k \pm eV/2)  $
and $f_{ab} (k) \equiv \frac{1}{2} \sum_{\pm}  \, \pm f(\epsilon_k \mp eV/2)  $ 
($f$ is the Fermi distribution).
Elastic current and noise are obtained in a straightforward manner using Eq.~\eqref{current}
and reproduce Landauer-B\"uttiker~\cite{blanter2000} formulas with the energy dependent transmission 
\[
{\cal T} (\epsilon ) = {\cal T}_0 + \sin 2 \delta_0 \, \frac{\alpha_1 \epsilon}{T_K} + (\alpha_2 \sin 2 \delta_0 + \alpha_1^2 \cos 2 \delta_0)
\left( \frac{\epsilon}{T_K} \right)^2,
\]
with ${\cal T}_0 = \sin^2 (\delta_0)$. The lowest order is given by (${\cal R}_0=1-{\cal T}_0$)
\begin{equation}\label{noise0}
S_0 = \frac{2 N e^2}{h} \left ( {\cal T}_0 {\cal R}_0 e V \coth\left( \frac{e V}{2 T} \right) + 2 T {\cal T}_0^2 \right).
\end{equation}
Corrections due to elastic terms $\sim \alpha_{1/2}$ can be written using functions ${\cal W}_1 = (T /2) ((eV)^2 + 4 (\pi T)^2/3 )$,
${\cal W}_2 = (e V/12) \coth ( eV/2 T) ( (eV)^2 + 4 (\pi T)^2)$,
${\cal A}_1 = \alpha_1^2 ( \cos 2 \delta_0 + 2 \sin^2 (2 \delta_0)-1)
+ \alpha_2 \sin 2 \delta_0 (1-\cos 2 \delta_0)$, and ${\cal A}_2 = \alpha_1^2 \cos 4 \delta_0 +\frac{\alpha_2}{2} \sin 4 \delta_0$,
\begin{equation}\label{noise1}
 \delta S_e = \frac{2 N e^2}{h \, T_K^2} \left( {\cal W}_1 {\cal A}_1 + {\cal W}_2 {\cal A}_2  \right).
\end{equation}

We now detail electron interactions within the Fermi liquid picture. 
As explained by Nozi\`eres~\cite{nozieres1974}, 
scattering off the frozen singlet results in electron phase shifts 
$\delta_\sigma(\epsilon,n_{\sigma'})$ depending on the 
energy $\epsilon$ and other spin ($\sigma'$) distributions
$n_{\sigma'}$.
Assuming analyticity, a low energy expansion in $\epsilon$ and $n_{\sigma'}$
can be written
\begin{equation}\label{expansion}
\delta_\sigma = \delta_{\rm el} (\epsilon) 
-  \frac{\sum_{\sigma'\ne\sigma, \epsilon'} (\phi_1 + \phi_2 \frac{\epsilon+\epsilon'}{2 T_K}) 
 \, n_{\sigma'} (\epsilon')}
{\nu T_K},
\end{equation}
with the phenomenological coefficients $\alpha_1$, $\alpha_2$ 
(see Eq.~\eqref{dephasing}), $\phi_1$, $\phi_2$.
As the Kondo resonance position is determined relatively to the Fermi energy,
these coefficients are not independent~\cite{nozieres1974}, one finds $\alpha_1 = (N-1) \phi_1$
and $\alpha_2 = \frac{(N-1)}{4} \, \phi_2$,
in accordance with conformal field theory arguments~\cite{affleck1993,lehur2007}.
Universality is recovered as $\alpha_1$ sets all other coefficients.

\begin{figure}
\includegraphics[width=5.5cm]{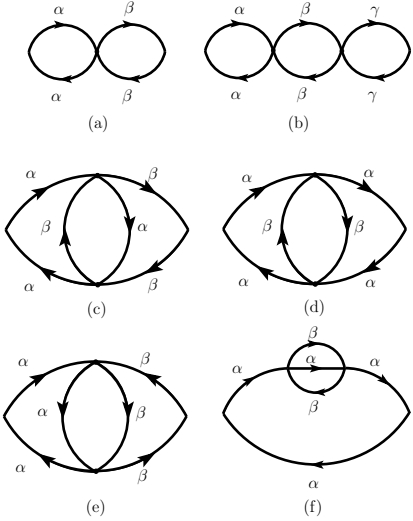}
\caption{\label{diag} Diagrams for the noise appearing in
the first and  second order expansions in the inelastic Hamiltonian Eq.~\eqref{interaction}.
Diagrams (c) and (d) give identical contributions.
For diagram (f), the three-lines bubble can alternatively dress the
bottom Green's function.  $\alpha,\beta,\gamma$ denote spins. }
\end{figure}

The inelastic part of Eq.~\eqref{expansion} is obtained from
\begin{equation}\label{interaction}
\begin{split}
H_{\rm int} = \frac{\phi_1}{\pi \nu^2 T_K} \sum_{\sigma < \sigma'} : b_{\sigma}^\dagger (0) b_{\sigma} (0)
b_{\sigma'}^\dagger (0) b_{\sigma'} (0) : \\[2mm]
+ \frac{\phi_2}{\pi \nu^2 T_K^2} \sum_{\sigma < \sigma',\{ k_i \}}
\frac{\epsilon_{k_1} + \epsilon_{k_2}}{2}  : b_{\sigma,k_1}^\dagger b_{\sigma,k_2}
b_{\sigma',k_3}^\dagger b_{\sigma',k_4}  :,
\end{split}
\end{equation}
where  $:\,\,:$ denotes normal order.
Inelastic corrections  are calculated perturbatively.
Since scattering states annihilated by $c_{L/R,k}$ are eigenstates of the Hamiltonian for
$H_{\rm int}=0$, they form a convenient basis for a perturbation calculation in $H_{\rm int}$
in the Keldysh framework.
The corresponding diagrams for the noise are shown Fig.\ref{diag}.
They are obtained by connecting current operator vertices to interaction four-point vertices.
Internal lines involve only $b$ fermions whereas external current vertices mix $a$ and $b$
fermions, see Eq.~\eqref{current}. Non-interacting Green's functions are combinations of $L/R$
thermal distributions, with ${\cal G}_{ a b} = {\cal G}_{b a}$, ${\cal G}_{ a a} = {\cal G}_{b b}$,
${\cal G}^{\alpha,\beta}_{ a b} (\omega, k) = 2 i \pi \delta(\omega-\epsilon_k) f_{ab}(k)$
for all Keldysh indices $(\alpha$,$\beta)$ and, for instance,
${\cal G}^{+-}_{b b} (\omega, k) = 2 i \pi \delta(\omega-\epsilon_k) f_{bb}(k)$,
${\cal G}^{-+}_{b b} (\omega, k) = 2 i \pi \delta(\omega-\epsilon_k) (f_{bb}(k)-1)$.
Consistency in the calculation requires
to stop at quadratic order in $max(e V,T)/T_K$ for inelastic corrections.
Therefore only $\delta_0$ is kept in Eq.~\eqref{dephasing} for second order diagrams  (b)-(f), while 
$\sim \alpha_1$ corrections are included in diagram (a). 
Moreover the second term in the Hamiltonian Eq.~\eqref{interaction} enters solely diagram (a).

The elastic part of Eq.~\eqref{expansion} is exactly included in the scattering states.
It is therefore  not necessary here to use
an elastic Hamiltonian~\cite{elastic} in contrast to  Refs.~\cite{sela2006,gogolin2006b,golub2006}.
We have nevertheless checked that the two approaches give coinciding results.
To write results in a concise way, we define the prefactor ${\cal S}_P \equiv (2 e^2/h)\, N (N-1) (\phi_1/T_K)^2$
and the functions ${\cal G}_1 = \frac{e V}{6} ( (eV)^2 + (\pi T)^2 )$, 
 ${\cal G}_2 = \frac{e V}{12} ( (eV)^2 + 4 (\pi T)^2 )$, 
 ${\cal G}_3 = T ( 5 (eV)^2 + \frac{8}{3} (\pi T)^2 )$, ${\cal F}_1 = \coth (e V /T)$,
${\cal F}_2 = \coth (e V/2 T)$. Collecting diagrams (c)-(f), we find the contribution
\begin{equation}\label{noise2}
\begin{split}
& \delta S_{i,1} = {\cal S}_P \Big[ {\cal G}_1 \Big( ( 2 \cos 4 \delta_0 +2) {\cal F}_1 - 2 \sin^2(2 \delta_0)
{\cal F}_2 \Big) \\[2mm] &+  {\cal G}_2 {\cal F}_2 \cos^2(2 \delta_0) + \frac{2}{3} \, \sin^2(2 \delta_0) \pi^2 T^3
+ {\cal G}_3  \sin^2(\delta_0) \cos 2 \delta_0 \Big].
\end{split}
\end{equation}
Diagram (b) gives 
\begin{equation}\label{noise3}
\delta S_{i,2} = 2 {\cal S}_P (N-1) {\cal T}_0 {\cal R}_0 (e V)^2 \left( \frac{e V}{2} \coth \left( \frac{e V}{2 T} \right)
+ T \right).
\end{equation}
 The last diagram (a) gives a mixed contribution with elastic and inelastic parts,
\begin{equation}\label{noise4}
\delta S_{m} =  -  {\cal S}_P \, (e V)^2 T \, {\cal T}_0
{\cal R}_0 \left( 4  \frac{\alpha_1}{\phi_1} + \sqrt{\frac{{\cal T}_0}{{\cal R}_0}} \frac{\phi_2}{\phi_1^2} \right).
\end{equation}
To summarize  the total noise reads
$S = S_0 + \delta S_e + \delta S_{i,1} + \delta S_{i,2} + \delta S_m$
where the different terms are given Eqs.~\eqref{noise0},\eqref{noise1},\eqref{noise2},\eqref{noise3},
\eqref{noise4}. This is the {\it central result} 
of this letter. 

At low voltage ($e V\ll T$), 
the Johnson-Nyquist relation is verified, {\it i.e.} $S = 4 T G$ where $G =  \partial I/\partial V|_{V=0}$ is the
linear conductance.
It is interesting to study more specifically the shot noise (at $T=0$).
In contrast with the SU(2) case, shot noise does not vanish at the Fermi liquid
fixed point. For $T_K \to \infty$, $S=S_0 = 2 (e V/h) e^2 N {\cal T}_0 {\cal R}_0$. 
Hence this term dominates at low energy, elastic and inelastic contributions
come only as corrections to $S_0$.
The inelastic term Eq.~\eqref{interaction} describes scattering of  $0$, $1$ or $2$
electrons from left scattering channel (L) to right (R) or vice versa~\cite{sela2006}.
At zero temperature, three processes are relevant: (i) (L,L) $\to$ (R,R), (ii)
(L,R) $\to$ (R,R), (iii) (L,L) $\to$ (R,L) with rates 
\[
\Gamma_1 =  N (N-1) \frac{e V}{h} \, \frac{\phi_1^2}{48}  \left( \frac{e V}{T_K} \right)^2,
\]
for (ii), (iii),
and $\Gamma_2 = 8 \Gamma_1$ for (i). Diagram (a) does not contribute at zero temperature.
Eq.~\eqref{noise2} gives
\begin{equation}\label{shot1}
\delta S_{{\rm i},1} =  8 \Gamma_1 (e^*)^2 + 2 \Gamma_2  [ (2 e^*)^2 - 8 {\cal R}_0 {\cal T}_0 e^2],
\end{equation}
where $e^* = e ( 1- 2 {\cal T}_0)$.
The first two terms can be identified as two uncorrelated backscattering events with effective
charges $e^*$ and $2 e^*$. This picture is in agreement with the inelastic current correction
that we find $\delta I_i = 4 \Gamma_1 e \cos 2 \delta_0 + \Gamma_2 2 e \cos 2 \delta_0$.
The third term  in Eq.~\eqref{shot1} is a reduction of the partition noise. It is balanced by
diagram (b) that adds a positive contribution to the partition noise~\cite{notepro}
\begin{equation}\label{shot2}
\delta S_{{\rm i},2} =  \frac{2 e^2}{h} \, N(N-1)^2 \left( \frac{ \phi_1}{T_K} \right)^2 {\cal T}_0 {\cal R}_0
(eV)^3
\end{equation}
Gathering this term and the last one of Eq.~\eqref{shot1}, we find an overall increase
of the partition noise due to Fermi liquid interactions. This can be interpreted as an 
increase in
the density of states of (L)electron-(R)hole pairs -involved in the partition noise- due
to repulsive electron interactions.

\begin{figure}
\includegraphics[width=5cm]{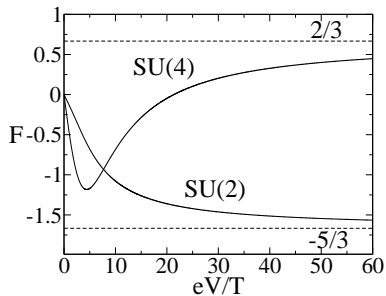}
\caption{\label{fig-fano} Generalized Fano factor $F$ from Eq.~\eqref{fano} as a function
of $e V/T$ (full lines) for SU(2) and SU(4). Dotted lines are shot noise limits.}
\end{figure} 

To analyse the effect of temperature, we define a generalized experimentally relevant Fano factor
\begin{equation}\label{fano}
F (e V/T) \equiv \frac{1}{2 e}
\dfrac{S(V,T)-S_0(V,T)-4 T \frac{\partial \delta I}{\partial V} (V,T)}{\delta I(V,T)},
\end{equation}
with $\delta I(V,T) = I(V,T) - I_0(V)$, $I_0 (V) = (N e^2 V /h ) {\cal T}_0 $ 
is the current for $T_K \to \infty$.
 $T_K$ disappears in Eq.~\eqref{fano} and $F$ 
becomes a universal function
of the ratio $e V/T$, shown Fig.~\ref{fig-fano} for SU(2) and SU(4).
In the shot noise limit, $F = -5/3$ for SU(2) in accordance with Refs.~\cite{sela2006,gogolin2006b},
and $F = 2/3$ for SU(4). $F=-1$ is obtained if only elastic terms are present for $N=2,4$.
For larger $N>4$, $F$ converges towards a universal increasing function where only elastic terms
contribute and with $F=1$ in the shot noise limit.
In the case of SU(4), noise is suppressed by elastic terms but enhanced by inelastic ones.
The overall effect is a noise increase and a positive $F$ -noise and current corrections are both
positive- in contrast with SU(2).
The convergence to the shot noise limit is quite slow, see Fig.~\ref{fig-fano}. For SU(4), $F$ can
even change sign, depending on the temperature, due to the negative 
$\delta S_m$, Eq.~\eqref{noise4}, competing with other terms at intermediate temperatures.
It clearly shows that temperature corrections are important in a large window of experimentally 
accessible parameters.

To conclude, we have determined the current noise through a quantum dot in the Kondo Fermi
liquid regime. The result holds for general SU(N) symmetry and arbitrary $V/T$.
We find in particular that temperature corrections are important for comparison with
experiments. For the SU(4) case, electron interactions are shown to enhance the noise
yielding a positive effective Fano factor in contrast with SU(2).
We stress that our approach is not restricted to the present problem but could be
applied to mesoscopic systems where elastic scattering is accompanied by
weak inelastic processes.

\acknowledgments
We thank T. Kontos, P. Simon, R. Combescot, and K. Le Hur  for fruitful discussions.
We thank especially  A.~A.~Clerk for pointing us the difference
in the evaluation of diagrams (c)-(d) and (e).



\newcommand{{{\PRB}}}{{{Phys. Rev. B}}}\newcommand{{{\PR}}}{{{Phys. Rev.}}}\newcommand{{{\PRA}}}{{{Phys. Rev. A}}}\newcommand{{{\PRL}}}{{{Phys. Rev. Lett}}}\newcommand{{{\NPB}}}{{{Nucl. Phys.}}}\newcommand{{{\RMP}}}{{{Rev. Mod. Phys.}}}\newcommand{{{\ADV}}}{{{Adv. Phys.}}}\newcommand{{{\EPJB}}}{{{Eur. Phys. J. B}}}\newcommand{{{\EPJD}}}{{{Eur. Phys. J. D}}}\newcommand{{{\JPSJ}}}{{{J. Phys. Soc. Jpn.}}}\newcommand{{{\JLTP}}}{{{J. Low Temp. Phys.}}}


\begin{thebibliography}{79}
\bibitem{elastic}\bibinfo{note}{this can be written $H_{\rm el} = - \frac{1}{2 \pi \nu T_K} \sum_{k,k',\sigma} ( \alpha_1 \xi_{k,k'}  + \alpha_2 \xi_{k,k'}^2) b^\dagger _{k,\sigma} b_{k',\sigma}$, with $\xi_{k,k'} \equiv = \hbar v_F (k+k')$}.
\end{thebibliography}
\end{document}